\documentclass[twocolumn,showpacs,preprintnumbers,amsmath,amssymb]{revtex4}

\usepackage{graphicx}
\usepackage{dcolumn}
\usepackage{bm}

\usepackage[utf8]{inputenc}
\DeclareUnicodeCharacter{200B}{}
\DeclareUnicodeCharacter{2009}{\,}
\DeclareUnicodeCharacter{2062}{}

\begin{document}

\preprint{APS/123-QED}

\title{Gapless behavior in a two-leg spin ladder with bond randomness}

\author{Yu Tominaga$^{1}$, Itsuki Shimamura$^{1}$, Takanori Kida$^{2}$, Masayuki Hagiwara$^{2}$, Koji Araki$^{3}$, Yuko Hosokoshi$^{1}$, Yoshiki Iwasaki$^{4,5}$, and Hironori Yamaguchi$^{1,5}$}

\affiliation{
$^1$Department of Physics, Osaka Metropolitan University, Osaka 599-8531, Japan\\
$^2$Center for Advanced High Magnetic Field Science (AHMF), Graduate School of Science, Osaka University, Osaka 560-0043, Japan\\
$^3$Department of Applied Physics, National Defense Academy, Kanagawa 239-8686, Japan\\
$^4$Department of Physics, College of Humanities and Sciences, Nihon University, Tokyo 156-8550, Japan\\
$^5$Innovative Quantum Material Center (IQMC), Osaka Metropolitan University, Osaka 599-8531, Japan
}


Second institution and/or address\\
This line break forced

\date{\today}

\begin{abstract}
We successfully synthesized [Cu$_2$(AcO)$_4$($p$-Py-V-$p$-F)$_2$]$\cdot$4CHCl$_3$, a verdazyl-based complex with a paddlewheel structure comprising two Cu atoms, which induces strong antiferromagnetic (AF) exchange interactions between Cu spins, generating a nonmagnetic singlet state at low temperatures.
Two primary exchange interactions between radical spins generate a spin-1/2 AF two-leg ladder.
In addition, two possible positional configurations of the F atom in the complex create four different overlap patterns of molecular orbitals, introducing bond randomness in the spin ladder.
The observed experimental behaviors, such as the Curie tail in the magnetic susceptibility and the gapless gradual increase in the magnetization curve, are attributed to a broad distribution of excitation energies and a few orphan spins in the random-singlet (RS) state that are stabilized by bond randomness.
The low-temperature specific heat exhibits a temperature dependence with $\propto 1/|{\rm{ln}}T|^3$, demonstrating the formation of the RS state in unfrustrated systems.
We also consider the effect of restricted patterns of exchange interactions and one-dimensional nature of the system on the RS state. 
\end{abstract}


\maketitle
\section{INTRODUCTION}
Spin ladders, which comprise coupled spin chains, are crucial for investigating low-dimensional quantum systems, bridging one-dimensional chains and two-dimensional lattices. 
These systems provide insights into phenomena driven by the interplay between dimensionality and quantum fluctuations, with potential relevance to high-temperature superconductivity~\cite{SL1,SL2}.
The spin-1/2 antiferromagnetic (AF) two-leg ladder, in particular, serves as a fundamental model, revealing the complex physics of ladder systems. 
In these systems, spins along the rungs form singlet pairs, leading to a gapped ground state. 
When a magnetic field is applied, the spin gap gradually closes and the system transitions into a Tomonaga-Luttinger liquid (TLL) phase---a gapless quantum critical state. 
Numerous spin ladder compounds have been experimentally identified and extensively studied, consistently confirming the predicted quantum behaviors, including the closing of the spin gap and emergence of the TLL phase~\cite{SL_exp1,SL_exp2,SL_exp3,SL_exp4}.

Randomness plays a crucial role in determining the physical properties of condensed matter systems. 
Even a small amount of disorder can notably alter the quantum behavior, leading to the emergence of novel quantum phases and critical phenomena. 
This disorder can have profound effects on various material properties, including transport phenomena, magnetic ordering, and quantum phase transitions.
The effect of randomness is particularly prominent in low-dimensional quantum spin systems, which are highly susceptible to quantum fluctuations and disorder. 
These systems give rise to the gapless quantum state known as the random singlet (RS) state.
In the last century, extensive theoretical studies have investigated the ground state of the 1D spin-1/2 chain with randomness, identifying it as the unfrustrated RS state~\cite{24,25,26,27}.
For the frustrated case, recent numerical studies on various spin-1/2 systems have shown that the introduction of randomness universally induces the RS state, referred to as the frustrated RS state~\cite{ram3,ram4,ram5,ram6,ram8}.
Moreover, a quantum phase transition between the unfrustrated and frustrated RS states has been identified, demonstrating that these two RS states exhibit qualitatively different low-temperature behavior~\cite{uematsu_JPSJ}.
For spin ladders, numerous studies have explored site randomness and its influence on quantum phases and critical phenomena~\cite{site1,site2,site3}. 
However, bond randomness, where disorder affects the interaction strength between spins, remains less studied. 
While theoretical predictions suggest that bond randomness in spin-1/2 AF two-leg ladders can significantly alter quantum phases and quasiparticle excitations~\cite{ram_ladder1, ram_ladder2, ram_ladder3}, experimental realizations of bond randomness are scarce and its effects have not been thoroughly investigated.

We developed a unique class of materials, verdazyl-based quantum organic materials (V-QOM), which offer exceptional flexibility in molecular design and precise control over spin interactions. 
Previous studies with V-QOM have demonstrated the realization of unconventional spin lattices that are yet to be generated in traditional inorganic materials~\cite{3Cl4FV, TCNQ_square, PF6, LM, SbF6}. 
By leveraging the properties of V-QOM, we successfully introduced randomness in exchange interactions by generating regioisomers that randomly align within the crystals. 
This led to the realization of a random singlet (RS) state in a spin-1/2 honeycomb lattice, marking a significant advancement in the understanding of quantum phenomena induced by bond randomness~\cite{random}. 
These developments represent a robust basis for exploring mechanisms underlying quantum criticality.


In this study, we synthesized [Cu$_2$(AcO)$_4$($p$-Py-V-$p$-F)$_2$]$\cdot$4CHCl$_3$ (AcO = CH$_3$COO$^{-1}$, $p$-Py-V-$p$-F = 3-(4-pyridinyl)-1-(4-fluorophenyl)-5-phenylverdazyl), a verdazyl-based complex. 
The four bridging acetates and two $p$-Py-V-$p$-F ligands formed a paddlewheel structure, where the strong antiferromagnetic (AF) exchange interactions between Cu spins created a spin-1/2 AF dimer, generating a nonmagnetic singlet state at low temperatures.
We identified two primary exchange interactions between the radical spins that generated the spin-1/2 AF two-leg ladder.
Two possible positional configurations for the F atom in the complex introduced bond randomness in the spin ladder, stabilizing the RS state.
Magnetic properties revealed AF gapless behavior with low paramagnetic contributions in the RS state.
The low-temperature specific heat exhibited the $1/|{\rm{ln}}T|^3$ behavior predicted for the unfrustrated RS state.  
We also examined the effect of restricted patterns of exchange interactions and  one-dimensional nature of the systems on the RS state.

\begin{figure*}[t]
\begin{center}
\includegraphics[width=38pc]{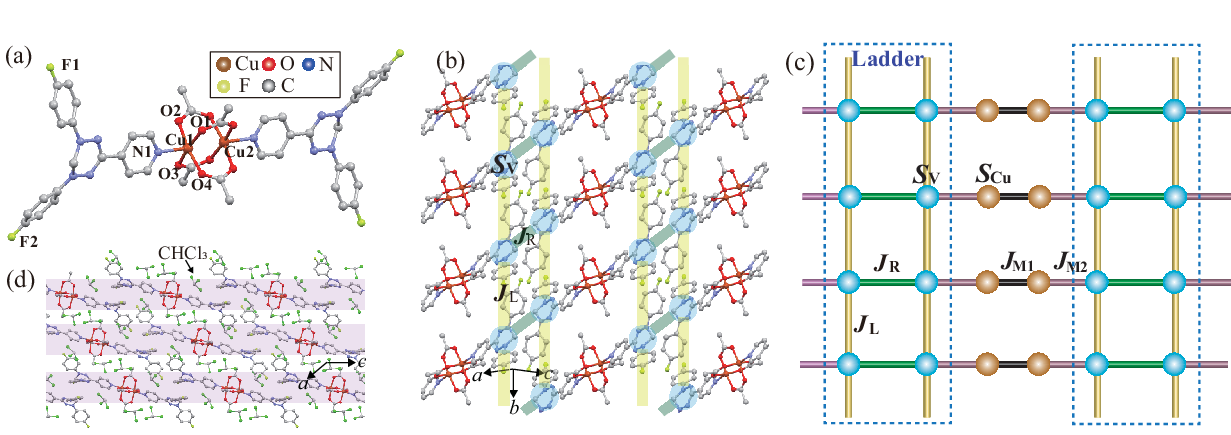}
\caption{(color online) (a) Molecular structure of [Cu$_2$(AcO)$_4$($p$-Py-V-$p$-F)$_2$], which causes intramolecular exchange interactions $J_{\rm{M1}}$ between the Cu spins and $J_{\rm{M2}}$ between the radical and Cu spins. 
The hydrogen atoms have been omitted for clarity. 
(b) The two-dimensional (2D) crystal structure and (c) corresponding spin-1/2 2D lattice with two different spin sites, $S_{\rm{V}}$ and $S_{\rm{Cu}}$.
The broken lines indicate the effective spin ladder.
(d) Crystal structure in the $ac$ plane. 
The thick lines represent the 2D spin planes.
}
\end{center}
\end{figure*}

\begin{figure}[t]
\begin{center}
\includegraphics[width=20pc]{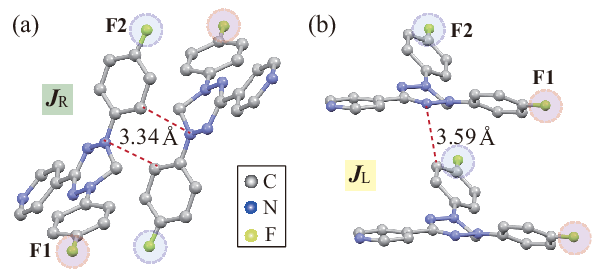}
\caption{(color online) Intermolecular radical pairs of [Cu$_2$(AcO)$_4$($p$-Py-V-$p$-F)$_2$]$\cdot$4CHCl$_3$ associated with exchange interactions (a) $J_{\rm{R}}$ and (b) $J_{\rm{L}}$.
Hydrogen atoms and Cu$_2$(AcO)$_4$ are omitted for clarity. 
The dashed lines indicate N-C short contacts.
}\label{f1}
\end{center}
\end{figure}

\section{EXPERIMENTAL}
We synthesized $p$-Py-V-$p$-F via the conventional procedure for producing the verdazyl radical~\cite{verd}.
A solution of Cu(AcO)$_2$$\cdot$H$_2$O (79.9 mg, 0.40 mmol) in 7 ml ethanol was slowly added to a solution of $p$-Py-V-$p$-F (132.9 mg, 0.40 mmol) in 8 ml of CH$_2$Cl$_2$ and stirred for 30 min. 
A dark-green crystalline solid of [Cu$_2$(AcO)$_4$($p$-Py-V-$p$-F)$_2$] was separated by filtration. 
Single crystals were obtained via recrystallization from CHCl$_3$ at 10 $^\circ$C.

The X-ray intensity data were collected using a Rigaku XtaLAB Synergy-S instrument.
The crystal structures was determined using a direct method using SIR2004 and refined using the SHELXL97 crystal structure refinement program.
Anisotropic and isotropic thermal parameters were employed for non-hydrogen and hydrogen atoms, respectively, during the structure refinement. 
The hydrogen atoms were positioned at their calculated ideal positions.
Magnetization measurements were conducted using a commercial SQUID magnetometer (MPMS, Quantum Design).
The experimental results were corrected by considering the diamagnetic contributions calculated using Pascal's method.
High-field magnetization in pulsed magnetic fields was measured using a non-destructive pulse magnet at AHMF, Osaka University.
Specific heat measurements were performed using a commercial calorimeter (PPMS, Quantum Design) employing a thermal relaxation method.
All the experiments utilized small, randomly oriented single crystals.

Molecular orbital (MO) calculations were performed using the UB3LYP method.
The basis sets are 6-31G (intermolecule) and 6-31G($d$, $p$) (intramolecule).
All calculations were performed using the GAUSSIAN09 software package.
The convergence criterion was set at 10$^{-8}$ hartrees.
We employed a conventional evaluation scheme to estimate the intermolecular exchange interactions in the molecular pairs~\cite{MOcal}. 

The quantum Monte Carlo (QMC) code is based on the directed loop algorithm in the stochastic series expansion representation~\cite{QMC2}. 
The calculations were performed for $N$ = 256 under the periodic boundary condition, where $N$ denotes the system size.
All calculations were carried out using the ALPS application~\cite{ALPS,ALPS3}.

\section{RESULTS}
\subsection{Crystal structure and spin model of the complex}
The crystallographic parameters of [Cu$_2$(AcO)$_4$($p$-Py-V-$p$-F)$_2$]$\cdot$4CHCl$_3$ are provided in Table I.
Figure 1(a) shows the molecular structure, wherein the verdazyl radical, $p$-Py-V, and Cu$^{2+}$ have a spin value of 1/2.
Table II lists the bond lengths and angles of the Cu atoms.
The four bridging acetates and two $p$-Py-V-$p$-F ligands form a distorted square pyramidal geometry around the Cu atoms---typical of Cu$^{2+}$ carboxylate dimers---which is known as the paddlewheel structure.
An inversion center located at the Cu$_2$ core, and the Cu–Cu distance (2.61 $\rm{\AA}$) is comparable with those of the other paddlewheel structures~\cite{Cu_dimer1,Cu_dimer2}. 
A Cu-Cu bond is not formed due to the lack of properly oriented orbitals for overlap, resulting in a five-coordinated square pyramid.
Two possible positional configurations are observed for the F atom (Fig. 1(a)), with rations of 0.786(9) and 0.214(9) for F1 and F2, respectively, as determined by X-ray single crystal analysis.
These two configurations introduced randomness in the MO overlap. 
Regarding the spin density distribution in the radicals, MO calculations showed that $\sim $61 ${\%}$ of the total spin density was localized on the central ring consisting of four N atoms, while each fluorophenyl ring directly attached to the central N atom contributed $\sim $16 ${\%}$ of the spin density.
Dominant exchange interactions were identified through MO calculations.
Regarding intramolecular interactions, the AF exchange interactions between the Cu spins in the paddlewheel structure and those between the radical spin and Cu spin were evaluated to be $J_{\rm{M1}}/k_{\rm{B}}$ = $572$ K and $J_{\rm{M2}}/k_{\rm{B}}$ = $4.9$ K, respectively, defined within the Heisenberg spin Hamiltonian, given by $\mathcal {H} = J_{n}{\sum^{}_{<i,j>}}\textbf{{\textit S}}_{i}{\cdot}\textbf{{\textit S}}_{j}$.
Given that MO calculations tend to overestimate the intramolecular interactions with transition metals, the actual values are expected to be smaller~\cite{Mn,octa}.
Regarding intermolecular interactions, two primary exchange interactions, $J_{\rm{R}}$ and $J_{\rm{L}}$, generated a spin-1/2 two-leg ladder along the $b$ axis, as depicted in Fig. 1(b).
Thus, the interactions $J_{\rm{M1}}$, $J_{\rm{M2}}$, $J_{\rm{R}}$, and $J_{\rm{L}}$ from a spin-1/2 two-dimensional (2D) lattice with two distinct spin sites, $S_{\rm{V}}$ and $S_{\rm{Cu}}$ (Fig. 1(c)).
Due to the presence of nonmagnetic CHCl$_3$ molecules between the 2D structures, exchange interactions between the spin lattices are expected to be weak enough to be neglected in this study (Fig. 1(d)).

We also evaluated the values of $J_{\rm{R}}$ and $J_{\rm{L}}$, which correspond to the rung and leg interactions in the spin ladder, respectively. 
The molecular pairs associated with $J_{\rm{R}}$ and $J_{\rm{L}}$ are related by inversion symmetry and translational symmetry along the $b$-axis, respectively, as depicted in Figs. 2(a) and 2(b).
Since the actual radicals have either F1 or F2, each overlap of the MOs associated with $J_{\rm{R}}$ and $J_{\rm{L}}$ has four radical pairing patterns with different F positions, i.e., F1-F1, F1-F2, F2-F1, and F2-F2. 
The possibilities of these pairing patterns and corresponding evaluations of $J_{\rm{R}}$ and $J_{\rm{L}}$ are summarized in Table III.
Notably, $J_{\rm{R}}$ strongly depends on the F position, and the pairs F1-F2 and F2-F1 are equivalent due to the inversion center between the radicals.
In contrast, $J_{\rm{L}}$ exhibits weak dependence on the F position, although the uniform bond is distorted by randomness in MO overlap.

\begin{table}
\caption{Crystallographic data of [Cu$_2$(AcO)$_4$($p$-Py-V-$p$-F)$_2$]$\cdot$4CHCl$_3$.}
\label{t1}
\begin{center}
\begin{tabular}{lc}
\hline
\hline 
Formula & C$_{50}$H$_{44}$Cu$_{2}$Cl$_{12}$F$_{2}$N$_{10}$O$_{8}$\\
Crystal system & Monoclinic \\
Space group & $P2_{1}/n$ \\
Temperature (K) & 100 \\
$a$ $(\rm{\AA})$ & 19.0001(5) \\
$b$ $(\rm{\AA})$ & 9.5444(2)  \\
$c$ $(\rm{\AA})$ & 19.0221(4)  \\
$\beta$ (degrees) & 114.183(3) \\
$V$ ($\rm{\AA}^3$) & 3146.84(12) \\
$Z$ & 2 \\
$D_{\rm{calc}}$ (g cm$^{-3}$) & 1.587\\
Total reflections & 3621 \\
Reflection used & 3064 \\
Parameters refined & 391 \\
$R$ [$I>2\sigma(I)$] & 0.0561  \\
$R_w$ [$I>2\sigma(I)$] & 0.1511 \\
Goodness of fit & 1.041 \\
CCDC & 2405110\\
\hline
\hline
\end{tabular}
\end{center}
\end{table}

\begin{table}
\caption{Bond lengths ($\rm{\AA}$) and angles ($^{\circ}$) related to the Cu atoms in [Cu$_2$(AcO)$_4$($p$-Py-V-$p$-F)$_2$]$\cdot$4CHCl$_3$.}
\label{t1}
\begin{center}
\begin{tabular}{cc@{\hspace{1.5cm}}cc}
\hline
\hline
Cu1--N1 & 2.16 & N1--Cu1--O1 & 95.6 \\
Cu1--O1 & 1.96 & O1--Cu1--O3 & 169.3 \\
Cu1--O2 & 1.98 & O3--Cu1--N1 & 95.0 \\
Cu1--O3 & 1.97 & O1--Cu1--O2 & 89.6 \\
Cu1--O4 & 1.96 & O2--Cu1--O3 & 89.0 \\
Cu1--Cu2 & 2.61& O3--Cu1--O4 & 90.4 \\
 &  & O4--Cu1--O1 & 89.1 \\
 &  & N1--Cu1--O2 & 94.5 \\
 &  & O2--Cu1--O4 & 169.3 \\
 &  & O4--Cu1--N1 & 96.2 \\
\hline
\hline
\end{tabular}
\end{center}
\end{table}

\begin{table}
\caption{Possible radical pair patterns with different F positions and the values of $J_{\rm{R}}$ and $J_{\rm{L}}$ evaluated from MO calculations.}
\label{t1}
\begin{center}
\begin{tabular}{c@{\hspace{1 cm}}c@{\hspace{1 cm}}c@{\hspace{1 cm}}c}
\hline
\hline
Pattern & Possibility & $J_{\rm{R}}/k_{\rm{B}}$ (K) & $J_{\rm{L}}/k_{\rm{B}}$ (K) \\
\hline
F1--F1 & 0.618 & 8.0 & 1.6 \\
F1--F2 & 0.168 & 5.8 & 1.6 \\
F2--F1 & 0.168 & 5.8 & 2.0 \\
F2--F2 & 0.046 & 3.8 & 2.0 \\
\hline
\hline
\end{tabular}
\end{center}
\end{table}

\begin{figure}[t]
\begin{center}
\includegraphics[width=20pc]{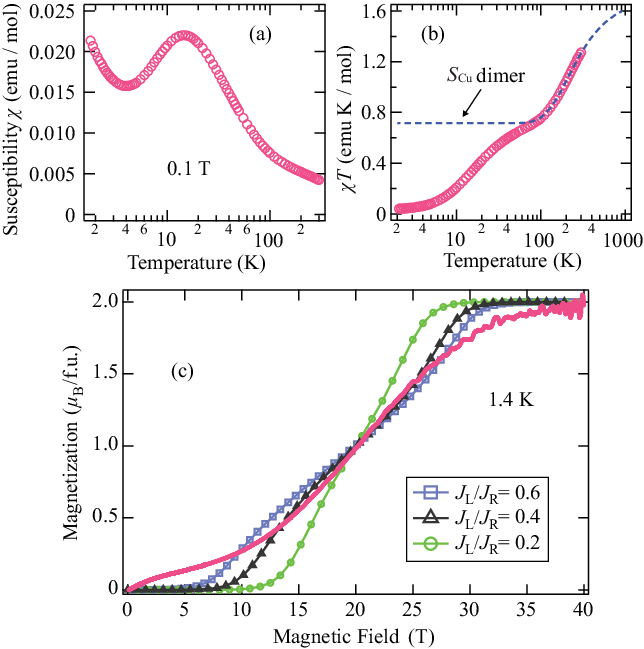}
\caption{(color online) Temperature dependence of (a) the magnetic susceptibility ($\chi$ = $M/H$) and (b) $\chi$$T$ of [Cu$_2$(AcO)$_4$($p$-Py-V-$p$-F)$_2$]$\cdot$4CHCl$_3$ at 0.1 T. 
The broken line represents the result calculated for a spin-1/2 AF dimer of $S_{\rm{Cu}}$ via $J_{\rm{M1}}$.
(c) The magnetization curve of [Cu$_2$(AcO)$_4$($p$-Py-V-$p$-F)$_2$]$\cdot$4CHCl$_3$ at 1.4 K under applied pulsed magnetic fields. 
The solid lines show calculated magnetization curves at 1.4 K for the spin-1/2 ladder with the representative values of $J_{\rm{L}}$/$J_{\rm{R}}$.
}\label{f1}
\end{center}
\end{figure}

\begin{figure}[t]
\begin{center}
\includegraphics[width=19pc]{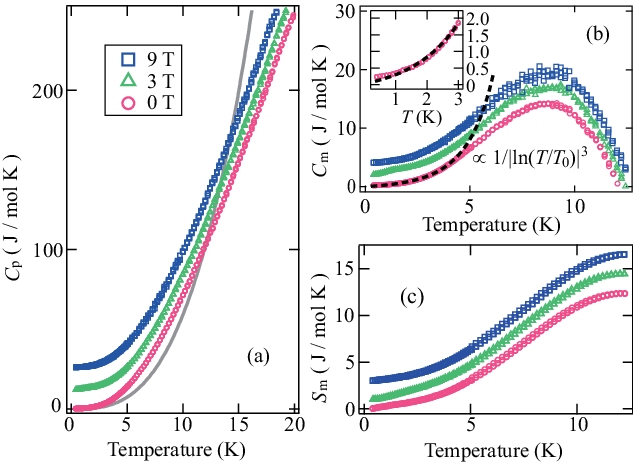}
\caption{(color online) 
(a) Temperature dependence of the specific heat $C_{\rm{p}}$ of [Cu$_2$(AcO)$_4$($p$-Py-V-$p$-F)$_2$]$\cdot$4CHCl$_3$.
The values at 3 and 9 T have been shifted up by 12 and 26 J/mol K, respectively.
The solid line indicates the lattice contribution by assuming Debye's $T^{3}$ law.
(b) Temperature dependence of the magnetic specific heat $C_{\rm{m}}$ evaluated by subtracting the lattice contribution.
The values at 3 and 9 T have been shifted up by 2 and 4 J/mol K, respectively.
The broken line indicates the $1/|{\rm{ln}}T/T_{\rm{0}}|^3$ behavior.
The inset shows the expansion of the low-temperature region at 0 T.
(c) Magnetic entropy $S_{\rm{m}}$ obtained by integrating $C_{\rm{m}}$/T. 
Here, we shifted the values at 3 and 9 T upward by 1 and 3 J/mol K, respectively.
}\label{f3}
\end{center}
\end{figure}

\begin{figure}[t]
\begin{center}
\includegraphics[width=17pc]{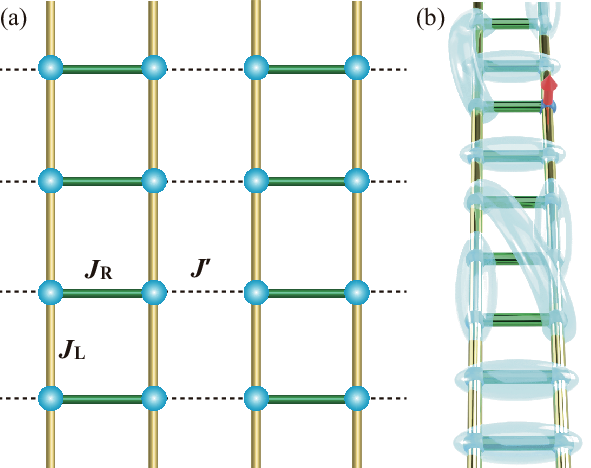}
\caption{(color online) 
(a) Two-leg spin ladder in [Cu$_2$(AcO)$_4$($p$-Py-V-$p$-F)$_2$]$\cdot$4CHCl$_3$. 
The effective inter-ladder interaction $J'$ occurs through the triplet excited states of the $S_{\rm{Cu}}$ dimer.
(b) RS state in the studied spin ladder. 
Entangled singlet dimers are indicated by ovals that cover two lattice sites. 
The dimers can be formed in a spatially random manner not only between neighboring sites but also between distant sites through higher-order interactions.
The red arrow represents an unpaired orphan spin.
}\label{f4}
\end{center}
\end{figure}

\subsection{Magnetic and thermodynamic properties}
Figure 3(a) shows the temperature dependence of the magnetic susceptibility ($\chi$) of the complex at 0.1 T, showing a broad peak at $\sim$14 K. 
The upturn below $\sim$4 K suggests paramagnetic contribution.
Assuming conventional paramagnetic behavior $C_{\rm{para}}$/$T$, where $C_{\rm{para}}$ is the Curie constant of paramagnetic spins, the paramagnetic contribution is found to be in the range of 3${\%}$--5${\%}$ of all spins.
This relatively large paramagnetic contribution could not be reduced by purification processes that have been effective in other verdazyl-based compounds.
The temperature dependence of $\chi$$T$, as shown in Fig. 3(b), reveals a two-step decrease as the temperature decreases.
The initial decrease, down to $\sim$100 K, is attributed to the formation of a spin-1/2 singlet dimer of $S_{\rm{Cu}}$ resulting from the strong AF interaction $J_{\rm{M1}}$. 
Further decrease at low temperatures reflects AF behavior originating from residual radical spins forming the spin-1/2 ladder.
We calculated the magnetic susceptibility of the spin-1/2 AF dimer of $S_{\rm{Cu}}$ and fitted it to the experimental $\chi$$T$ data above $\sim$100 K.
The calculated value was shifted up by 0.72 emu K/mol, corresponding to a Curie constant for spin-1/2 systems with $g$ = 2 and a radical purity of 96${\%}$, when accounting for the contribution of $S_{\rm{V}}$ (Fig. 3(b)).
The experimental behavior was then explained using $J_{\rm{M1}}/k_{\rm{B}}$ = 446(3) K and $g$ = 2.33(1), consistent with previously reported values for the paddlewheel structure~\cite{43,44,45,46}.


Figures 3(c) shows the magnetization curve of the complex under a pulsed magnetic field applied at 1.4 K.
The magnetization increases gradually up to $\sim$10 T, showing gapless behavior, and then rises sharply toward $\sim$30 T. 
The value of $\sim$2.0 $\mu_{\rm{B}}$/f.u. suggests full polarization of $S_{\rm{V}}$ with an isotropic $g$ value of 2.0, corresponding to the saturation of the spin ladder.
A slight paramagnetic contribution observed below 5 T, resembling the Brillouin function and consistent with the low-temperature upturn of $\chi$.
In typical spin ladders forming a gapped singlet state, the energy gap decreases under applied magnetic field, leading to a quantum phase transition to the TLL phase with a sharp increase in the magnetization~\cite{ladderMH1, ladderMH2, ladderMH3}.
However, the observed magnetization behavior deviates from this quantum critical behavior.


The temperature dependence of specific heat $C_{\rm{p}}$ is shown in Fig. 4(a). 
There is no peak showing phase transition toward magnetic long-range order. 
We assumed the lattice contribution of $C_{\rm{p}}$ to be ${\alpha}T^3$, corresponding to Debye's $T^{3}$ law for the low-temperature region, which is confirmed to be effective below $\sim$10 K in verdazyl-based compounds~\cite{3Cl4FV,ladderMH3}.
We then evaluated the magnetic specific heat $C_{\rm{m}}$ by subtracting the lattice contribution of $C_{\rm{p}}$, as shown in Fig. 4(b). 
The magnetic contributions contain inaccuracies in higher temperature regions, leading to the negative values above $\sim$12 K. 
We determined the value of ${\alpha}$ such that the magnetic entropy $S_{\rm{m}}$, which was obtained by integrating $C_{\rm{m}}$/$T$, exhibited asymptotic behavior near the broad peak temperature in $\chi$ below which the AF correlation was expected to emerge appreciably.
The obtained value ${\alpha}$ =0.059, corresponding to a Debye temperature of 32 K, was close to that of similar verdazyl-based complex~\cite{morotaMn}. 
Furthermore, the entropy change in the low-temperature region was close to the total magnetic entropy associated with a spin ladder composed of $S_{\rm{V}}$ (11.5; 2$R$ln2).
The $C_{\rm{m}}$ exhibited a broad peak, which was almost unchanged by the application of magnetic fields with magnitudes of up to 9 T.
This magnetic-field-insensitive behavior demonstrated that the broad peak did not arise from a Schottky-type behavior related to an energy gap; rather, it resulted from the development of AF correlations in the spin-1/2 ladder.
Considering the energy scale observed in the magnetization curve, the change in entropy induced by the application of a magnetic field is expected to be significant above $\sim$10 T, leading to a corresponding change in the specific heat values.




\section{Discussion}
We also examined the ground state of the spin model.
The Cu spins coupled by strong AF $J_{\rm{M1}}$ interactions formed a nonmagnetic singlet dimer at low temperatures.
Consequently, an exchange interaction $J'$ through the triplet excited states of the $J_{\rm{M1}}$ dimer was expected to occur between the radical spins, as depicted in Fig. 5(a). 
Assuming $T$ $\ll$ $J_{\rm{M1}}/k_{\rm{B}}$, a second-order perturbation treatment of the exchange interaction between $S_{\rm{V}}$ and $S_{\rm{Cu}}$---i.e., $J_{\rm{M2}}$---yielded $J' = J_{\rm{M2}}^{2}/2J_{\rm{M1}}$~\cite{Cu1, b26Cl2V}.
Considering $J_{\rm{M1}}/k_{\rm{B}}$ = 446 K determined from the magnetization analysis and $J_{\rm{M2}}/k_{\rm{B}}$ = 4.9 K determined from MO calculations, we estimated the value of $J'$ to be ${\textless}$0.1 K, which was negligible at our experimental temperatures.
Hence, the spin model in the low-temperature region can be regarded as a spin-1/2 AF two-leg ladder composed of $S_{\rm{V}}$.

Further, we discuss the effects of bond randomness in the prepared spin ladder.
Bond randomness is expected to stabilize the RS state, wherein singlet dimers of varying strengths are formed in a spatially random manner~\cite{24}. 
Because the exchange interactions are randomly distributed in the RS state, the binding energies of the singlet dimers have a broad distribution, showing gapless behavior. 
For magnetic susceptibility, Curie-like diverging components appear in the low-temperature region, reflecting paramagnetic contribution from some unpaired orphan spins~\cite{ram3,ram5}. 
For the prepared system, we observed the Curie tail in $\chi$, which suggests the existence of orphan spins specific to the RS state.
The evaluated amount of the paramagnetic contribution is indeed larger than those evaluated in other verdazyl-based compounds, which originate only from the lattice defects~\cite{3Cl4FV,TCNQ_square}.
Because of the lattice symmetry of the spin ladder, a rung singlet alone, i.e., $J_{\rm{R}}$ ${\textgreater}$ $J_{\rm{L}}$, cannot generate the orphan spins.
Therefore, the actual values of $J_{\rm{R}}$ and $J_{\rm{L}}$ apparently differ somewhat from the MO evaluations presented in Table III, which suggests that $J_{\rm{R}}$ ${\textless}$ $J_{\rm{L}}$ partially exists.
In this case, $J_{\rm{L}}$ allows the formation of singlet dimers, which yields the orphan spins, as illustrated in Fig. 5(b). 


For the magnetization curve, the RS state is predicted to exhibit a near-linear behavior, reflecting the wide distribution of excitation energies~\cite{ram3}.
However, our experimental result for the magnetization curve reveal a gapless but nonlinear behavior. 
We attribute this difference to the effect of randomness, which is limited by the characteristics of the present system.
Bond randomness is limited to three $J_{\rm{R}}$ patterns and two $J_{\rm{L}}$ patterns, and the single dimension of the lattice restricts the spatial distribution.
Consequently, although the excited state is gapless, the density of states depends strongly on the probability of existence of each exchange interaction.
The observed large increase in the magnetization above $\sim$10 T is considered to be due to the $J_{\rm{R}}$ interaction (F1-F1), which has the highest probability of existence.

To examine accurate energy scale of the present model, which strongly reflects the values of $J_{\rm{R}}$ (F1-F1), we calculated the magnetization curves  by considering $J_{\rm{L}}$/$J_{\rm{R}}$. 
Figure 3(b) shows the calculated results for representative values of $J_{\rm{L}}$/$J_{\rm{R}}$, demonstrating the difference from the experimental behavior. 
The magnetization curve for each $J_{\rm{L}}$/$J_{\rm{R}}$ was scaled to have the same energy scale by adjusting the value of $J_{\rm{R}}$.
As a result, identical intermediate fields toward saturation were obtained.
Consequently, the values of $J_{\rm{R}}/k_{\rm{B}}$ were determined as 24 K ($J_{\rm{L}}$/$J_{\rm{R}}$ = 0.2), 21 K ($J_{\rm{L}}$/$J_{\rm{R}}$ = 0.4), and 18 K ($J_{\rm{L}}$/$J_{\rm{R}}$ = 0.6).
Although quantitative comparison with the experimental results is unreliable due to the randomness effect, we can roughly estimate the actual value of $J_{\rm{R}}$ (F1-F1) to be $\sim$20 K, indicating the underestimation of the MO evaluations 
A relatively large difference from the MO evaluation may be related to the presence of the randomness in the molecular packing.


For the specific heat, numerical studies predict that a qualitative difference between unfrustrated and frustrated RS states appears in the low-temperature region~\cite{uematsu_JPSJ}.
The unfrustrated RS state, which corresponds to the present model, exhibits a specific heat behavior proportional to $1/|{\rm{ln}}T/T_{\rm{0}}|^3$, while the frustrated RS state follows a $T$-linear behavior. 
Experimentally, the magnetic specific heat shows a clear nonlinear behavior and is well reproduced by the $1/|{\rm{ln}}T/T_{\rm{0}}|^3$ fit for $T{\textless}$ 4 K, as shown in Fig. 4(b).
Since the low-temperature specific heat appears to be qualitatively equivalent even in magnetic fields, this behavior is considered to be robust against an applied magnetic field. 
The characteristic energy scale of the RS state, $T_{\rm{0}}$, is evaluated to be approximately 11 K, which is in close agreement with the energy scale observed in the magnetization data.





\section{Summary}
In this research, we synthesized a verdazyl-based complex: [Cu$_2$(AcO)$_4$($p$-Py-V-$p$-F)$_2$]$\cdot$4CHCl$_3$.
The four bridging acetates and two $p$-Py-V-$p$-F ligands formed a distorted square pyramidal geometry around the Cu atoms, creating a paddlewheel structure.
Strong AF exchange interactions between Cu spins generated a spin-1/2 AF dimer, stabilizing the nonmagnetic singlet state below $\sim$100 K.
We identified two primary exchange interactions between the radical spins, $J_{\rm{R}}$ and $J_{\rm{L}}$, which generated a spin-
1/2 AF two-leg ladder.
Two possible positional configurations of the F atom lead to four different MO overlap configurations associated with $J_{\rm{R}}$ and $J_{\rm{L}}$, introducing bond randomness in the spin ladder.
The magnetic susceptibility indicated a broad peak with a Curie tail in the low-temperature region, indicating AF correlations and low paramagnetic contribution.
The magnetization curve exhibited a gapless gradual increase up to $\sim$10 T, followed by a rapid rise toward spin-1/2 ladder saturation.
The specific heat showed a broad, magnetic-field-insensitive peak, indicating the development of AF correlations, rather than a Schottky peak related to an energy gap.
Furthermore, in the low-temperature region, the specific heat exhibited the $1/|{\rm{ln}}T|^3$ behavior predicted for the unfrustrated RS state.
The observed experimental behaviors of the system were attributed to a wide distribution of excitation energies, with a few orphan spins in the RS state.
In addition, the restricted patterns of exchange interactions and the one-dimensional nature of the system were considered to induce some deviations from the expected theoretical RS behavior. 
These findings provide insight into the effects of bond randomness on the quantum behavior of spin ladders. 
We propose a one-dimensional spin model of V-QOM with bond randomness.
This development is expected to inspire further research into the quantum phenomena arising from the interplay of quantum fluctuations and randomness.

\begin{acknowledgments}
This research was partly supported by KAKENHI (Grants No. 23K13065 and No. 23H01127) and the Iwatani Naoji Foundation.
A part of this work was performed under the interuniversity cooperative research program of the joint-research program of ISSP, the University of Tokyo.
\end{acknowledgments}


\end{document}